\documentclass[superscriptaddress,twocolumn,nobibnotes,nofootinbib,notitlepage,amsmath,amsfonts]{revtex4-1}
\usepackage{graphicx}
\usepackage{amsmath}
\usepackage{amsfonts}
\usepackage{amssymb}

\newcommand{\sciexp}[2]{{#1}\ensuremath{\,\times\,10^{#2}}}

\newcommand{\mfp}{{m\!f\:\!\!p}}

\newcommand{\tsp}{\hspace{3mm}}

\newcommand{\PU}{Department of Astrophysical Sciences, Princeton University, Princeton, NJ 08544, USA}
\newcommand{\PPPL}{Princeton Plasma Physics Laboratory, Princeton, NJ 08543, USA}
\newcommand{\LPP}{Laboratoire de Physique des Plasmas, \'{E}cole Polytechnique, Paris 75252, France}
\newcommand{\LLE}{Laboratory for Laser Energetics, University of Rochester, Rochester, New York 14623, USA}
\newcommand{\UNH}{Space Science Center, University of New Hampshire, Durham, New Hampshire 03824, USA}

\begin{document}

\title{Kinetic simulation of magnetic field generation and collisionless shock formation in expanding laboratory plasmas}
\author{W.~Fox}
\email{wfox@pppl.gov}
\affiliation{\PPPL}
\affiliation{\PU}

\author{J.~Matteucci}
\affiliation{\PU}

\author{C.~Moissard}
\affiliation{\LPP}

\author{D. B. Schaeffer}
\affiliation{\PU}

\author{A.~Bhattacharjee}
\affiliation{\PPPL}
\affiliation{\PU}

\author{K.~Germaschewski}
\affiliation{\UNH}

\author{S.X. Hu}
\affiliation{\LLE}


\begin{abstract} 

Recent laboratory experiments with laser-produced plasmas
have observed and studied a number of fundamental 
physical processes relevant to magnetized astrophysical 
plasmas,
including magnetic reconnection, collisionless shocks, and magnetic
field generation by Weibel instability,
opening up new experimental platforms for laboratory astrophysics.
We develop a fully kinetic simulation model for
first-principles simulation of these systems
including the dynamics of 
magnetic fields---magnetic field generation by the Biermann battery effect or
Weibel instability; advection by the ion flow, Hall effect, and Nernst effect;
and destruction of the field by dissipative mechanisms.
Key dimensionless parameters describing the system are derived for scaling
between kinetic simulation, recent experiments, and astrophysical plasmas.
First, simulations are presented which model Biermann battery 
magnetic field generation in plasmas expanding from a thin target.
Ablation of two neighboring plumes leads to the formation
of a current sheet as the opposing Biermann-generated fields collide,
modeling recent laser-driven magnetic reconnection experiments.
Second, we simulate recent experiments on collisionless
magnetized shock generation, by expanding a piston plasma
into a pre-magnetized ambient plasma.  For parameters 
considered, the Biermann effect generates additional magnetic fields
in the curved shock front and thereby increases shock particle reflection.
Both cases show the importance of
kinetic processes in the interaction of plasmas with magnetic fields, 
and open opportunities to benchmark these important
processes through comparison of theory and experiments.
\end{abstract}

\maketitle

\section{Introduction}

Laboratory experiments provide valuable platforms to test
fundamental physics behind astrophysical plasmas.
Studying and comparing the common physical
processes between systems with such differing varying spatial and temporal scales is
possible because of universality of the processes, 
embodied in scaling similarity of the underlying equations 
\citep{RyutovApJ2000, BuckinghamPhysRev1914}.
The recent generation of laser facilities allows
observation and study of a variety of processes in magnetized
plasmas relevant to plasma astrophysics.
Expanding laser-produced plasmas readily
produce fast supersonic flows, and
the plasmas can be magnetized either through
self-magnetization, or can be expanded into
plasmas pre-magnetized with an externally
applied field.
Recent laser facilities couple sufficient energy into the plasma
to obtain high-temperature, collisionless regimes while
simultaneously being large enough for scale separation between global scales and
kinetic scales.
The subsequent dynamics of the magnetized plasmas are of fundamental interest
and have broad relevance to some of the
most energetic processes in astrophysical plasmas, including
shock formation, magnetic reconnection, kinetic instabilities,
and particle energization. 

Strong, MG-scale magnetic fields can be generated in 
expanding laser plasmas by the Biermann battery effect \citep[e.g.][]{StamperPRL1971, StamperPRL1978, YatesPRL1982, PetrassoPRL2009},
or by kinetic instabilities such as Weibel instability \citep{FoxPRL2013, HuntingtonNatPhys2015}
and Rayleigh-Taylor instability \citep{ManuelPRL2012,GaoPRL2012}.  
These processes all have astrophysical analogues,
allowing experimental investigation of the generation of
magnetic fields and subsequent dynamics 
of magnetized plasmas in scaled laboratory astrophysics experiments.
The Biermann battery effect has been proposed to provide
primordial, seed magnetic fields in the cosmos, which 
can then be subsequently amplified by turbulence \citep{KulsrudApJ1997}.
The ion-Weibel instability is important for 
generating a turbulent magnetic field in interpenetrating plasmas,
mediating collisionless astrophysical shocks in unmagnetized 
or weakly magnetized regimes \citep{MedvedevApJ1999,KatoApJL2008}.
The interaction of magnetized flows can also steepen
into magnetized shocks \citep{SmithScience1975, BambaApJ2003}.
Collision of magnetized plasmas can drive magnetic reconnection, which
is an important mechanism for converting magnetic energy back to
particle flows, heat, and energized particles \citep{OierosetPRL2002, DrakeApJ2010, YamadaRMP2010}. The development
of both reconnection and shock experiments on laser facilities
will potentially allow comparison of the
efficiency of these two processes for energizing particles 
in astrophysics.

The evolution of the magnetic field in a plasma 
is determined from the Generalized Ohm's law, via Faraday's law.
In laser-produced plasmas, the Ohm's law can be written as \citep{EpperleinPoF1986}:
\begin{equation}
\mathbf{E} = \eta \mathbf{j} - \mathbf{v} \times \mathbf{B}
+ \frac{ \mathbf{j \times B}}{ne} - \mathbf{v}_T \times \mathbf{B}
- \frac{\nabla p_e}{ne} - \frac{\nabla \cdot \mathbf{\Pi}_e} {ne}.
\label{Eq_Ohm}
\end{equation}
The terms on the RHS enumerate the effects which evolve the field at scales ranging from
global to kinetic scales, which 
include resistive dissipation ($\eta \mathbf{j}$); advection of the magnetic field by the bulk fluid 
($\mathbf{v} \times \mathbf{B}$), the Hall effect ($\mathbf{j \times B}$), and the Nernst effect ($\mathbf{v_T \times B}$), where $\mathbf{v_T}$ is proportional to the heat flux; the
pressure term ($\nabla p_e$),  which allows electrothermal magnetic field generation; and momentum
transport embodied in the pressure tensor ($\nabla \cdot \mathbf{\Pi_e}$).  
The electron pressure can lead to magnetic field generation in
regions where the temperature and density gradients are not collinear
 (leading to finite $\nabla \times \mathbf{E}$), 
in what is often called the Biermann battery effect \citep{KulsrudApJ1997}.
The Nernst effect \citep{HainesPPCF1986}, which is advection of
magnetic field by the heat flux, can be important in semi-collisional regimes and 
results from the v$^{-3}$ dependence of the
Coulomb collision frequency; intuitively, it can be understood from the idea that the
magnetic field remains frozen to a population of ``hot'' electrons but is
allowed to diffuse across the cold compensating return current.
Finally, momentum transport, i.e. off-diagonal pressure tensor ($\nabla \cdot \mathbf{\Pi}$),
is often important for breaking magnetic field lines
in magnetic reconnection current sheets in collisionless regimes \citep{BirnJGR2001}.

Modeling the dynamics of energy exchange between
the magnetic field, plasma flows, and energized particles 
in these systems is a computational grand challenge, because 
the relevant processes (magnetic reconnection and shocks)
couple global scales with plasma kinetic 
scales  (the ion and electron skin depths) characteristic of reconnection current sheets or shock ramps, 
yielding a challenging, multi-scale kinetic plasma problem.
Fully kinetic simulations are important tools for 
intrinsically kinetic processes, especially
for Weibel instability and for collisionless shocks, where
particle distributions develop strong counter-streaming components
significantly deviant from a thermal description.
As another example, during magnetic reconnection in collisionless and weakly-collisional regimes, 
the field-line breaking is typically mediated by momentum transport,
a kinetic effect resulting from meandering electron orbits in the current sheet.
A final example is in strongly-driven, semi-collisional regimes, 
where the heat flux can advect the magnetic fields via the 
Nernst effect \citep{HainesPPCF1986, KhoPRL1985, RidgersPRL2008}.
Challenges remain to correctly model the Nernst effect when the heat-flux
becomes non-local and deviates from classical predictions.
While interesting ongoing progress has aimed
to identify closures for including all these processes within fluid
simulations, including for magnetic reconnection \citep{LiuPRE2017, NgPoP2017}
and Biermann battery field generation proximate to shocks \citep{GrazianiApJ2015},
for the foreseeable future fully kinetic particle simulations will play
a vital role in simulating these processes.

Particle-in-cell (PIC) methods solve the kinetic Vlasov-Maxwell
system using quasi-particles to represent particle phase space.
The quasi-particles evolve according to the equations of motion for the particles,
interacting with the electromagnetic fields on a mesh; the fields in turn evolve 
according to Maxwell's equations, with current sources determined
self-consistently by the particles.
While highly computationally intensive, modern implementations of
the PIC technique scale well to the largest supercomputers.   Here we 
use the PSC code \citep{GermaschewskiJCP2016} and 
develop techniques to simulate large volumes of laser-plasma with PIC simulation.
We note that particle simulations have long been used for high-intensity relativistic
short-pulse (ps- and fs-class) laser-plasma interaction 
\citep[e.g.][]{WilksPRL1992}; here we present how these
techniques are generalized to study much larger volumes 
(presently cm$^2$ in 2-D or mm$^{3}$ in 3-D) and longer time scales (ns),
where it is not practical to resolve the laser wavelength.

In this paper we develop a fully kinetic ablation model for 
first-principles and end-to-end simulation of 
recent laser-plasma experiments on magnetic reconnection 
\citep{NilsonPRL2006, LiPRL2007b,ZhongNature2010,FikselPRL2014, RosenbergPRL2015}, Weibel instability
\citep{FoxPRL2013, HuntingtonNatPhys2015} and 
collisionless magnetized shocks \citep{NiemannGRL2014, SchaefferPRL2017}.
In the first part of the paper, we 
simulate Biermann battery magnetic field generation and evolution from spatially
localized heating in a thin target.
This enables simulation of multiple colliding plumes, in which the plume
collison compresses the opposing Biermann fields into a current sheet
drives magnetic reconnection \citep{NilsonPRL2006, LiPRL2007b,ZhongNature2010,RosenbergPRL2015}.
Previous simulations of reconnection in these plasmas
have largely relied on simplified models and geometries \citep{FoxPRL2011, FoxPoP2012b, TotoricaPRL2016},
and, while many physics insights into reconnection in these systems
have come from these simulations, they are also limited for comparison
with experiment because they require many 
initial parameters to be put in by hand.
The model we describe here allows end-to-end 
simulation of these experiments, and we 
present simulations of the 2-D evolution of colliding plumes up through 
the time of current sheet formation.
Capturing the field generation and plume expansion has significant physical implications,
as the  plasma parameters which determine the
magnetic reconnection regime (such as density in the reconnection layer,
velocity driving the inflow, and local magnetic fields) are
ultimately determined by the evolution of the plasmas up through
the current sheet formation process.  
It also enables full 3-D simulations of
magnetic reconnection in these systems, which will be reported separately \citep{Matteucci2017}.

In the second half of this paper, we simulate the evolution of
ablated plasma into a pre-existing, magnetized background plasma, as has been
studied in recent experiments on strongly-driven magnetic reconnection \citep{FikselPRL2014},
and formation of collisionless magnetized shocks \citep{SchaefferPRL2017}.
The development of this platform will allow for a number of studies in 
high-Mach number shocks ($M \gtrsim 5$) 
including particle acceleration and particle heating by shocks,
and shock dynamics including reformation and instabilities.
The simulations played an important role in the
development of this experimental platform
by showing how the system behaves for a range of driving parameters,
building intuition about the physics of the available experimental regimes.
In the case of magnetized shocks, the simulations
also predicted observables such as the development of 
``double density jumps'' in the collisionless shock front.
Here we present the details of this simulation model,
and present large-scale 2-D simulations which show
physics associated with the higher-dimensionality expanding shocks, including
Biermann battery magnetic field
within the shock front.  For the parameters
considered quantitatively modifies the shock by increasing particle reflection.
Weibel instability is also observed upstream and downstream of the shock \citep{MatsumotoScience2015}
consistent with instability of the counterstreaming ion populations generated 
by these high-Mach number shocks.

\section{Simulation Model}

Fully kinetic PIC simulations are conducted to model the plasma evolution and magnetic field
generation in plasma rapidly expanding from a thin, high-density target,
enabling direct kinetic modeling of recent laser-plasma experiments.
The key point of the model is to match heating and plasma conditions
to separate, off-line radiation hydrodynamic (RH) simulations using the DRACO code \citep{HuPoP2013}, 
providing rigorous and
well-benchmarked parameters from which to drive the kinetic simulations.
The scheme of matching the relevant dimensionless parameters
from the simulation model to experiments is discussed below.
For experimental fidelity, 
the model includes particle collisions which allows collisional effects
such as magnetic field diffusion due to resistivity and magnetic field
advection by the Nernst effect.  Particle simulations are conducted using the 
PSC code \citep{GermaschewskiJCP2016}, which can run on recent
leadership scale computing facilities with good scaling and includes GPU support.

The kinetic ablation model applies a volumetric heating operator to a thin target
to heat the target and drive an expanding ablation plume.
The target region is simultaneously kept at constant density $n_s$ by continually
adding particles, mimicking the infinite reservoir provided by a 
solid density target.  The heated region expands as an ablation flow into the
neighboring low density plasma or vacuum.  
In our model, the sonic point occurs at 
 $n_{ab}$ = 0.4 $n_s$; 
therefore our model includes plasma density up to about 2.5 times the critical
density.
The plumes are observed to obtain a 
quasi-isothermal electron temperature $T_{ab}$, which depends
on the magnitude of the heating.
In the simulation, we adjust the heating to provide a desired $T_{ab}$.

Off-line DRACO RH simulations 
using reported laser parameters provide input 
to drive the kinetic PSC simulations, by the following procedure.
First, the DRACO simulations are conducted and analyzed.
These simulations generate 
an expanding plasma plume profile in which a characteristic ``knee" forms in the density at the sonic point of the ablative plasma flow.
This ablation point defines the ablation density $n_{ab}$ 
which is matched to the PIC model.
Secondly, outside the knee, the plasma is typically nearly isothermal,
and this provides an ablation temperature $T_{ab}$.
These provide the dimensional parameters for driving a scaled PSC simulation.
We find that by matching $n_{ab}$ and $T_{ab}$ in PSC,
the remaining hydro-dynamic evolution outside the 
ablation point is reasonably well-matched between the two simulations.
We note again PSC does not independently predict
$n_{ab}$ and $T_{ab}$ because it does not have a laser ray-trace energy-deposition model; 
this is the driving input from 
DRACO simulations.  Secondly, the PSC simulations do not include many features of 
the RH simulations, such as acceleration of the solid target by ablation pressure and
outward motion of the ablation surface.

\begin{figure}
\includegraphics{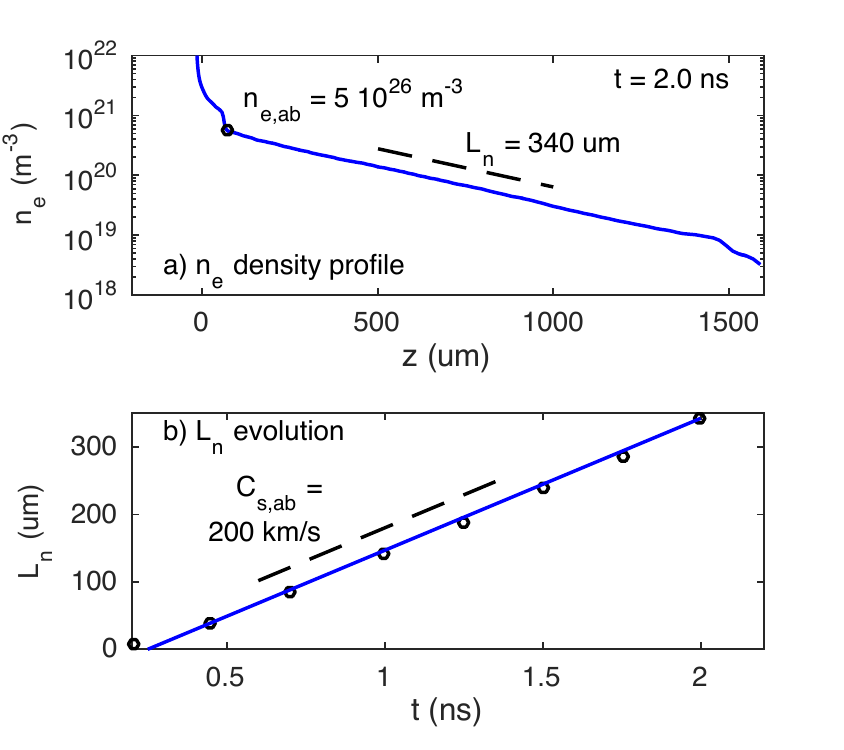}
\caption{Expanding plasma evolution predicted by DRACO, demonstrating measurement of
parameters to drive PSC simulations.  (a) $n_e$ profile off the target, showing
identification of $n_{ab}$ and the density scale length $L_n$ (b)
Evolution of the scale length.  In this DRACO simulation,
the density evolution $\langle\dot{L_n}\rangle$ agrees
quantitatively with the ablation sound speed $C_{s,ab} \approx$~\sciexp{2.0}{5}~m/s.}
\label{Fig_DRACO_Evol}
\end{figure}

\subsection{Dimensionless parameters}

Despite operating at vastly different 
length and time scales, results from laboratory experiments can be applied 
to understanding astrophysical phenomena, through scaling laws.
This is furnished by the nature of the underlying equations,
where it can be shown that systems are \textit{similar} 
such that the results can be scaled between systems,
provided that the relevant dimensionless parameters are matched \citep{RyutovApJ2000}.
The same set of considerations is necessary for numerical simulations as well,
as the latter are often conducted in dimensionless unit systems.
What is interesting is these considerations can allow
new simulation techniques to be applied, such as explicit PIC simulations as developed here,
by serving as a navigational aid in determining which 
parameters are most important to match, and which
need only be matched ``in regime''.
This is important because explicit PIC is a powerful technique, but also cannot match
every dimensionless parameter, especially those involving the
separation of electron and ion scales, due to computational cost.

To begin with, it is useful to determine the ``ion-scale'' parameters for this
family of systems.  We do this, assume that the plasma
dynamics depends on the following parameters: a
charateristic density and temperature 
$n_0$ and $T_0$, the ion mass $M_i$, ion charge $Ze$, $\mu_0$, 
and a spatial scale $R$, for example the heating radius.
Then by Buckingham's Pi theorem \citep{BuckinghamPhysRev1914}, given the 4
dimensional quantities (length, time, mass, charge), these 6 dimensional parameters
collapse to a much smaller space of only 2 dimensionless parameters:
and $R / d_{i0}$ and $n d_{i0}^3$.
The first parameter defines the scale separation between
the global scale and ion-kinetic scale and is fundamental for
this whole class of problems.
The second parameter $n d_{i0}^3$ is usually much larger than unity and 
is related to the particle discreteness and collisionality, 
and will return as a collisionality parameter in the discussion below.
The collisionality can be related to both the plasma
viscosity and magnetic diffusivity, as is commonly considered
in scaling analysis of the MHD equations for laboratory astrophysics \citep{RyutovApJ2000}.

As discussed above, $n_{ab}$ and $T_{ab}$ are
two important parameters describing the coronal plasma.
These two parameters form the
basis for scaling the length and time scales of the coronal plasma via the ``ion-scale'' parameters.
We define the ion skin depth $d_{i0} = (M_i / n_{ab} Ze^2 \mu_0)^{1/2} $ based on $n_{ab}$
and the characteristic ion species.  This provides the
fundamental length unit of the simulations.
Second, we define the ablation sound speed, $C_{s,ab} = (ZT_{ab}/M_i)^{1/2}$.
Together these define a characteristic 
ablation timescale $t_d = d_{i0} / C_{s,ab}$.
The scaling behavior enforces that the solutions for the
various physical quantities can be written
as a function of the dimensionless parameters and scaled coordinates, 
e.g. for the density,
$n = n_{ab} f( x/d_{i0}, t/t_d )$.
The classical 1-D ``ablation flow'' 
solution \citep{MannheimerPoF1982}, $n = n_{ab} \exp \left[- (x/C_s t) \right]$
exactly accords with this scaling behavior,
since $C_s = d_{i0}/t_d$.

The parameters also define a fundamental magnetic field scale $B_0 = (\mu_0 n_{ab} T_{ab})^{1/2}$.
Conveniently, in these units, $t \omega_{ci} = (B/B_0)\,(t/t_d)$,
and the local plasma $\beta = 0.5 (n/n_{ab}) (T/T_{ab}) (B/B_0)^{-2}$.
The fundamental unit of magnetic vector potential, $A_0 = d_{i0} B_0 = (1/e) (T_{ab} M_i/Ze)^{1/2}$,
is the integrated field (flux) sufficient to force ions at the 
sound speed to gyrate.
In the case with pre-applied magnetic fields, the upstream magnetic field
is parameterized by a further dimensionless parameter $B_{up} / B_{0}$.
The fundamental electric field is $E_0 = C_s B_0 = T_{ab} / d_{i0}$.
These scales are used below to convert simulation results to
physical units.

Explicit PIC simulations require two additional electron-scale parameters,
$Z m_e / M_i$, the electron-ion mass ratio,
and $T_{ab} / m_e c^2$, which describes the speed of light in the system.
These parameters describe the scale
separation of electron-ion physics and electrostatic phenomena,
as $d_{e} / d_{i}$ = $(Z m_e / M_i)^{1/2}$, and $\lambda_D / d_{e0} = (T_{ab} / m_e c^2)^{1/2}$,
where $\lambda_{D}$ is the Debye length.
A common technique in explicit PIC simulations (of magnetic reconnection and shocks)
is to compress these parameters compared to their physical values
so that they are matched in regime rather than exactly 
\citep[e.g.][]{BirnJGR2001, FoxPRL2011, FoxPoP2017, KatoApJL2008, MatsumotoScience2015}.
This allows important physical insights to be obtained from fully kinetic simulation
within reasonable computational cost, and tests can be 
conducted to verify the convergence of the
simulation results with respect to the reduced parameters. 

Finally, collisional processes in the plasma may be important to include
for experimental fidelity, by allowing thermalization of the distribution functions, 
or dissipation processes such as magnetic field diffusion.  In PSC, Coulomb collisions are modeled using the 
Takizuka-Abe binary collision model \citep{GermaschewskiJCP2016},
which is essentially equivalent to the Landau collision operator.
This introduces an additional time-scale, the
electron-ion collision time $\tau_{ei} = \nu_{ei}^{-1}$.
In the PSC code, $\nu_{ei}$ is a free parameter; therefore, it
generates an additional dimensionless parameter to match to the physical system.
We match the collisionality via the dimensionless parameter 
$\nu_{ei0} / \omega_{ce0} = d_{e0} / \lambda_{mfp0}$,
as this determines the magnetization character for
electrons in the Braginskii transport.
Here $\nu_{ei0}$ and $\lambda_{\mfp 0} = \sqrt{T_{ab}/m_e} / \nu_{ei0}$ are the 
electron collision frequency and mean-free path evaluated at $n_{ab}$ and $T_{ab}$,
and $\omega_{ce0}$ is the electron gyro-frequency at $B=B_0$.
Matching this dimensionless parameter 
also obtains the correct collisional diffusivity of the magnetic field.  
To show this, we 
calculate a magnetic Reynold's number based on sonic flows,
$R_M \equiv \mu_0 L C_s / \eta$, where $L$ is the system size and
$\eta$ the collisional resistivity.
  With a small amount of algebra,
$R_M = (L/d_{i}) \cdot (\lambda_{\mfp 0} / d_{e0})$, which
indicates that matching $L/d_{i0}$ as an ion-scale parameter,
and matching electron collisionality according to $\lambda_{\mfp 0} / d_{e0}$,
obtains the correct magnetic Reynold's number.
We do note that, following this scheme, the electron collisionality connected
to global time or length scales,
$L / \lambda_{\mfp 0}$, or $\nu_{ei0} t_d$ 
are only matched quantitatively if the physical mass ratio is also matched.  
This impacts, for example, studies of non-local transport, where the
electron mass dependence of the results would need to be carefully checked.

To summarize, in the kinetic ablation model,
dimensional analysis has reduced the problem to 
a small number of dimensionless parameters: $R/d_{i0}$
reflecting the scale separation between
the global and ion-kinetic scale;  $\lambda_{\mfp 0} / d_{e0}$ reflecting the collisionality;  
and the electron-scale dimensionless parameters $Zm_e/M_i$
and $m_e c^2 / T_{ab}$.  Additional applied magnetic 
fields are parameterized through $B_{up} / B_0 = B_{up} / (\mu_0 n_{ab} T_{ab})^{1/2}$.

\subsection{Particle deposition and heating}

As discussed above, kinetic ablation simulations maintain a high-density thin target
region, which is topped up to mimic the reservoir of plasma provided
by a solid-density target.  During the simulation, particles are topped up in the target region
using an implicit method
\begin{equation}
n \to n + \frac{n_s - n}{1 + \tau_p / \Delta t_p} \qquad \mathrm{if}\; n < n_s,
\end{equation}
where $\tau_p$ is a characteristic time to refill particles, and $\Delta t_p$ is the
time between calls to the particle deposition code.  For a wide range of 
$\tau_p / \Delta t_p$ the density is kept close to $n_s$. 
Particles are added at
a specified cold target temperature in appropriate electron-ion ratios to maintain a charge-neutral plasma.

Meanwhile, electrons are heated in the target by random kicks to the momentum vectors,
\begin{equation}
p_i \to p_i + \left(m_e \,H \, \Delta t_h \right)^{1/2} r
\end{equation}
where $r$ is a gaussian random variable with unity variance, $\Delta t_h$ is the time between calls to
the heating routine, and $H(x,y,z)$ is a spatially-dependent heating rate (eV/s).
$\Delta t_h$ reflects sub-cycling of the heating operator
compared to the particle push timestep, for efficiency.
The magnitude of $H$ is adjusted to obtain a desired $T_{ab}$.
A spatial profile can be applied to $H$, for example gaussian and super-gaussian
shapes characteristic of laser foci, for localized heating and Biermann-battery field generation.

\begin{figure}
   \centering
   \includegraphics{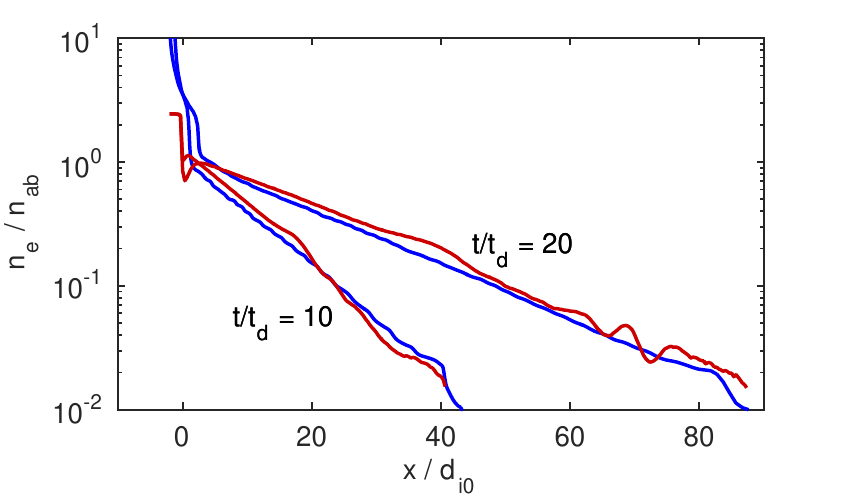}   
   \caption{Comparison of density profiles from DRACO (blue) and PSC (red), at two times.}
   \label{Fig_draco_psc}
\end{figure}

Despite its simplicity we find this model gives a good agreement
for coronal plasma evolution compared to radiation-hydro simulation
and analytic theory.  Figure~\ref{Fig_draco_psc} shows
a comparison of the plasma density between 
PSC and DRACO at two scaled times, showing the
agreement.  (For parameters, see Section~\ref{SectionShock}.)

\subsection{Further numerical details}

The plasma evolves from a thin high-density target held
at density $n_s$ by the top-up method described above, with typical thickness $3 d_{i0}$.
Plasma expands into a background plasma at density $n_{bg} \ll n_{ab}$.  For some
systems, such as for the shock studies, the background density is an important parameter to
match to experiment.  The background plasma and target are also initialized at 
a cold plasma temperature $T_c$.  In physical experiments $T_c$ is room temperature; here
we simply require $T_c \ll T_{ab}$.

PSC simulations are conducted on a uniform grid with grid spacing $\Delta x$ and $\Delta z$.
We find that there are restrictions on the magnitude of $\Delta x$ and $\Delta z$,
so that Debye-scale numerical heating does not play a role on the time scale of the simulation.
The most stringent location for numerical instability is in the initially cold high-density region.
There the instability criterion is based on the Debye length calculated with
$T_c$ and $n_s$, and we find that for the PSC second-order particle shape, $\Delta x /\lambda_{D,c} < 50$
is required to avoid
significant numerical heating on the simulation 
time scale, where $\lambda_{D,c} / d_{e0} = (n_{ab}/n_s)^{1/2} (T_c / m_e c^2)^{1/2}$

Finally, PSC simulations use macro-particles which represent
a certain volume of phase space, and we typically find that 50 particles per cell
representing $n_{s}$ is sufficient to well-resolve the physics under study,
though we typically use 200 or more in the 2-D simulations presented here.

\section{Magnetic field generation and evolution in laser-driven reconnection experiments}

The Biermann battery effect 
is important for magnetic field generation in both laser experiments and astrophysics.  
Non-collinear gradients of density and temperature lead to volumes with
net circulating EMF which can drive currents and generate a magnetic field.  
The Biermann battery has also been proposed as a mechanism in astrophysics to generate primordial magnetic
fields, predominantly near shocks, where the strong localized entropy production produces the 
sharp gradients for a strong 
thermoelectric effect \citep{KulsrudApJ1997}.  In laser-plasma interaction, the effect
is to self-magnetize the plasma, with a toroidal magnetic field wrapping around the laser focus.
This was been observed and studied in a number of early experiments
\citep{StamperPRL1971, StamperPRL1978, YatesPRL1982}, 
with recent resurgence driven by new measurement techniques based on proton radiography
\citep{PetrassoPRL2009, WillingalePRL2010, GaoPRL2015}, and use of multiple plasmas to 
drive magnetic reconnection between neighboring plumes at a small separation
\citep{NilsonPRL2006, LiPRL2007b, RosenbergPRL2015}.  

The experiments use pairs of laser-produced plumes from a flat
target. The Biermann effect generates a strong
magnetic field of order megagauss (MG), which forms a toroidal ribbon
wrapping around the bubble.  If multiple bubbles are created at small
separation, the bubbles expand into one another, squeezing the
opposing magnetic fields together and driving reconnection.
Laser-driven magnetic reconnection experiments have the possibility to significantly
advance understanding of laboratory
reconnection by, first, obtaining data on the reconnection behavior of large systems 
(measured by the ratio of the system size
to fundamental plasma scales such as the ion inertial length $d_i$).  To date, the majority of 
dedicated laboratory experiments \citep{YamadaRMP2010} have been conducted  at moderate system
sizes, where two-fluid 
effects have been demonstrated to be important for obtaining fast reconnection \citep{RenPRL2005}. 
In large systems it is is proposed that the tearing or plasmoid instability can break the current into a hierarchy of meso-scale
current sheets undergoing fast reconnection.  
Second, these experiments can measure 
the efficiency of particle acceleration by reconnection, again in large systems.  
Significant particle energization
is inferred to occur in many astrophysical environments, and magnetic reconnection
has been proposed to be the driver in many cases, including 
the Earth's magnetotail \citep{ZenitaniApJ2001, OierosetPRL2002}, 
solar flares \citep{LinSolPhys1976}, the Crab nebula and pulsar \citep{UzdenskyApJLett2011},  and extragalactic jets \citep{LarrabeeApJ2003}.  

Previous simulations have considered the reconnection in parameter
regimes of HEDP plasmas, and with model geometries of colliding bubbles
to assess reconnection mechanisms and particle acceleration.
Simulations have demonstrated the role of flux pile-up 
\citep{FoxPRL2011}, the breakup of the current sheet 
into multiple islands at large system size \citep{FoxPoP2012b},
and have documented mechanisms for generating electron jets
and accelerating particles \citep{LuNJP2014, LuNJP2016, TotoricaPRL2016, FoxPoP2017}.
Recent kinetic simulations have also studied Biermann-battery magnetic
field generation in collisionless plasmas with model initial conditions
 \cite{SchoefflerPRL2014}.

Here we present a kinetic simulation
of the full development of the current sheet
between colliding plasmas,
including the initial Biermann-battery 
field generation, advection and collision of the fields,
which thin down and compress into 
a current sheet where the magnetic fields reverse over a narrow region.
Through analysis of the Ohm's law we quantitatively assess the
various roles of the Biermann effect for generating field and subsequent
advection by the plasma flow, Hall effect, and Nernst effect.
The results show a supersonic advection of the magnetic field
by the Hall effect which has not been previously identified.
This work forms the basis for 
comparison of field generation and advection with detailed experiments
and complementary simulation techniques such as radiation-hydrodynamics 
which includes magnetic field effects \citep{GaoPRL2015, LanciaPRL2014}.
Full 3-D simulations which include reconnection will be 
presented in a followup paper.

We first briefly review the Biermann battery magnetic field generation mechanism.
To understand the mechanism of 
B-field generation, 
we first re-write the Ohm's law, keeping the plasma currents to make the causation more transparent, 
and ignoring collisions for the time being
\begin{equation}
(m_e/ne^2) \, \frac{d}{dt} \mathbf{j} = \nabla p_e/ne + \mathbf{E}.
\end{equation}
Now, consider a volume in the plasma with finite $\nabla p_e / ne$, with  $\nabla p_e$ non-parallel to $\nabla n$, and 
consider a loop in the plane defined by the vectors 
 $\nabla n \times \nabla T$.
If these two vectors are not everywhere parallel, then $n$ is not a function
of $T$, and integration of $\nabla p_e / ne$ around the loop is in general non-zero;
indeed by Stoke's theorem, $\oint (\nabla p_e / ne) \cdot d\mathbf{s} = 
\int (1/n e) (\nabla n_e \times \nabla T_e) \cdot d\mathbf{A}$, for a given path $d{s}$ following the
boundary of the surface $\mathbf{dA}$.
This ``battery" produces a net electromotive force (EMF) around the loop.
By itself this would drive a circulating current 
which would be extremely large due to the small electron mass.  Note that even if 
large electrostatic fields in the plasma arise (indeed also driven by the pressure gradient),
they integrate to zero around the loop.
Therefore, absent inductive electric fields, a large circulating current would be driven.  
However, the increasing circulating current creates an increasing magnetic field,
which induces an electric field by Faraday's law, which can
balance the Biermann EMF.
From this balance one obtains $\partial \mathbf{B}/ \partial t = \nabla \times (\nabla p_e/ne)$,
which is the standard result.
The only subtlety in the final equation is that the causal role 
of the electrons currents in driving the magnetic field 
is hidden.
Mathematically this is because in the Ohm's law, 
$\mathbf{j}$ is finite, but $m_e \mathbf{j}$ is small compared to 
$\mathbf{E}_{\mathrm ind}$ and $\nabla p_e/ne$ due to the small electron mass.
(With a little more analysis, the precise 
criterion for ignoring the electron inertia term is for
is for density and temperature
gradient scales larger than the electron skin depth, in which case the
energy in Biermann magnetic fields dominates the kinetic energy in electron flows.)

We analyze the development and evolution of the magnetic field
using the generalized Ohm's law implemented in the PSC code,
which takes a slightly different form than Eq.~\ref{Eq_Ohm},
\begin{equation} 
\mathbf{E} = - \left(\mathbf{v \times B} \right) 
  + \frac{1}{n_e e} \left( \mathbf{j \times B} \right) - \frac{\nabla p_e}{n_e e} - \frac{1}{n_e e} \nabla \cdot \Pi
  - \frac{1}{n_e e} \mathbf{R}_{ei}.
\end{equation}
The terms on the RHS are: the ion flow term;  Hall term;  electron pressure term, 
which can drive the Biermann effect; electron pressure tensor, which embodies momentum
transport; and collisional momentum exchange.  
Here we separate the pressure scalar and pressure tensor from the total stress tensor, by
defining the pressure tensor as trace-less.
The collisional momentum exchange $\mathbf{R}_{ei}$ includes both resistive diffusion and
the Nernst-effect (via the thermal force).  It
is obtained by diagnosing the total momentum exchange between electrons and ions 
during the collision subroutine.  This provides a rigorous measure of the momentum exchange and
confirmation of total momentum conservation, but does not allow ready separation of the diffusion and Nernst effects,
which requires extra analysis.

\subsection{Simulation setup and parameters}

Using the model described above we simulate the
 magnetic field generation and evolution of recent laser plasma experiments on magnetic reconnection.
The experiments were conducted
at Vulcan and Shenguang-II laser facilities.
(The experiments on OMEGA of \citet{LiPRL2007b} and \citet{RosenbergPRL2015} are very large in
($L / d_{i0}$) and will be considered in future work.)
Both experiments coupled several hundred J of laser energy onto thin metal targets with laser intensities
near 10$^{15}$~W/cm$^2$, producing high temperature ($T \sim$~1 keV), and high-density plasmas ($n \sim 10^{25}$~m$^{-3}$).

DRACO RH simulations were conducted to predict the plasma temperature $T_{ab}$
and density $n_{ab}$ to determine the parameters for PSC simulations
using the scheme above.  The simulations used quoted laser intensities $I \sim 10^{15}$~W/cm$^2$,
laser-incidence angles (45 degrees in both cases), and then Aluminum foils as the target material.
Table~\ref{TableShockExp} summarizes the results of these calculations
and presents the resulting dimensional 
parameters ($n_{ab}, T_{ab}, t_d, B_0$) for converting the PIC results
back to physical units.

\begin{table}
\centering
\caption{Parameters for Vulcan \citep{NilsonPRL2006} 
and SG-II \citep{ZhongNature2010,DongPRL2012} reconnection experiments}
\begin{tabular}{lrr}
\tableline\tableline
& Vulcan  & SG-II  \\

Input: \\
\tsp Ions & Al$^{13+}$ & Al$^{13+}$ \\
\tsp $n_{ab}$ (m$^{-3}$, DRACO) & \sciexp{5}{26} & \sciexp{4.5}{27} \\
\tsp $T_{ab}$ (eV, DRACO) & 1500 & 2000 \\
\tsp heating radius $R_H$  ($\mu$m)& 50 & 100 \\
\tsp Separation  $2L$ ($\mu$m) & 400 & 400 \\
\\

Dimensional: \\
\tsp $d_{e0}$ ($\mu$m) & 0.24 &  0.08 \\
\tsp $d_{i0}$ ($\mu$m) & 14 & 4.8 \\
\tsp $C_{s0}$ (m/s) & \sciexp{2.7}{5} & \sciexp{3.2}{5} \\
\tsp $t_d = d_{i0}/C_{s0}$ (ps) & 52 & 15 \\
\tsp $B_0 = (\mu_0 n_{ab} T_{ab})^{1/2}$ (T) & 380 & 1300 \\
\tsp $B_0 d_{i0}$ (T-mm) & 5 & 6 \\

\tsp $\omega_{ce0}$ (ps$^{-1}$) & 70 & 230 \\
\tsp $\nu_{ei0}$ (ps$^{-1}$) & 2.7 & 14 \\
\\
Dimensionless: \\
\tsp $R_H / d_{i0}$ & 2 & 15 \\
\tsp $L / d_{i0}$ & 12 & 40  \\
\tsp $\lambda_{\mfp 0} / d_{e0}$ & 25 & 15 \\

\end{tabular}
\label{TableReconExp}
\end{table}

\begin{table}
\centering
\caption{PSC Simulations for Vulcan \citep{NilsonPRL2006} and SG-II \citep{ZhongNature2010,DongPRL2012} reconnection experiments}
\begin{tabular}{lrr}

\tableline\tableline
& Vulcan  & SG-II  \\

Initialization: \\
\tsp $R_H / d_{i0}$ & 2 & 12 \\
\tsp $L / d_{i0}$ & 12 & 40  \\
\tsp $\lambda_{\mfp 0} / d_{e0}$ & 30 & 20 \\
\\
Results: \\
\tsp $t_{coll} / t_d$ & 5--10 & 15--20 \\
\tsp Max $\Psi$ / $B_0 d_{i0}$ & 3 & 3 \\
\end{tabular}
\label{TableReconSim}
\end{table}

\subsection{Simulation Results}

\begin{figure*}
\centering
\includegraphics[width=\linewidth]{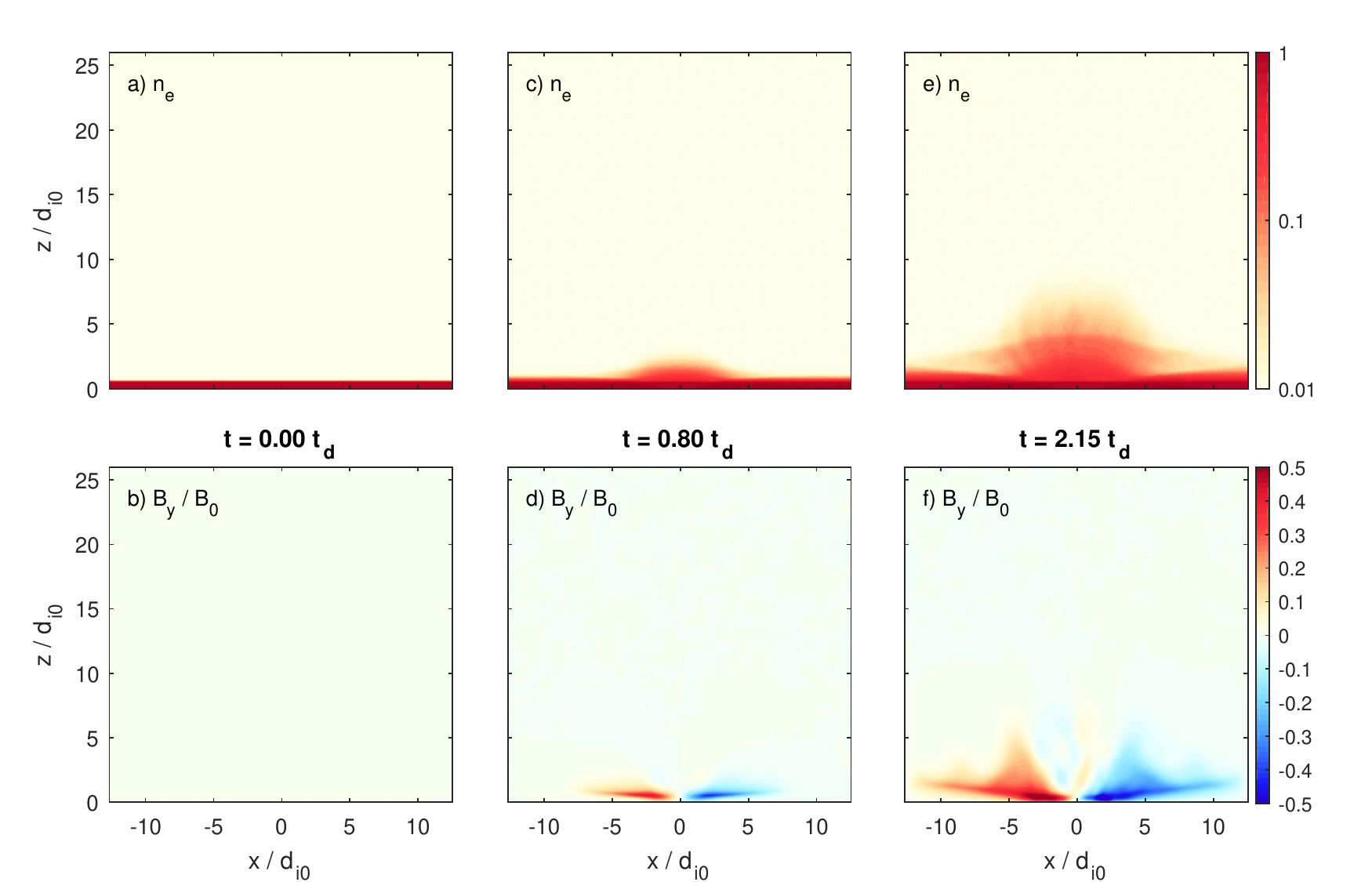}
\caption{Evolution of plasma density and magnetic field in a simulation model of the
experiments of Ref.~\citep{NilsonPRL2006}.  \label{Fig_evol}}
\end{figure*}

These PSC simulations represent the first kinetic simulations of Biermann battery 
magnetic field generation in plasmas with parameters
and geometry directly matched to these recent laser experiments.
The simulations further show the advection processes governing 
the magnetic field evolution, and resulting magnetic field compression
and current sheet formation when two neighboring plumes collide.
The simulations also reveal novel physics which has not be previously
reported, such as the role of the Hall effect for fast magnetic 
field advection over the target surface.  
Table~\ref{TableReconSim} summarizes the dimensionless parameters used for the
simulations as well as numerical details, and typical results such as
peak magnetic field strengths, and time for collision of a significant 
magnetic field for two plasmas initially separated by the distance $2L$.

Figure~\ref{Fig_evol} shows the plasma density and magnetic field evolution at three characteristic
times of the simulation modeling the experiments of Ref.~\citep{NilsonPRL2006}.
The left panel shows the initial condition with a thin high-density target surrounded
by a low-density ambient plasma, and zero magnetic field.
This plasma is heated with a gaussian heating profile with $1/e$ radius of $2 d_{i0}$.
The plume quickly reaches the ablation temperature $T_{ab}$ within the
heated region and expands. 
The evolution from $t= 0$ to $t = 2.15 t_d$ shows the generation of
Biermann magnetic fields near the target surface, produced in the
regions with non-collinear density and temperature gradients in the laser-heating
area, which subsequently expand vertically as well as radially outward.

Notably, peak magnetic fields up to 0.3 $B_0$ are rapidly generated, which correspond to
magnetic fields of order 1~MG, making use of the scalings of Table~\ref{TableReconExp}.
Second, a leading edge of 
magnetic field has been generated and advected approximately 10~$d_{i0} \approx
200$~um in $t = 2 t_d \approx 100$~ps, for magnetic advection rate of approximately 
5 times the sound speed $C_{s0}$, or roughly \sciexp{2}{6}m/s.
This fast expansion speed corresponds to a small leading edge of
magnetic field, and is qualitatively (but not quantitatively) consistent
with an extremely fast expansion inferred in some experiments at up to
\sciexp{8}{6}m/s \citep{WillingalePRL2010}.
We note that this leading edge of the magnetic field is fairly small
compared to the total magnetic flux, which is carried over a longer timescale of 4--5~$t_d$,
corresponding to speeds near 2--2.5~$C_{s0}$.
Line-integrated magnetic fields $\int B_y dz$ of order $3 B_0 d_{i0} \approx 15$ T-mm, 
(or 150 MG-um, in a common unit scheme) are obtained, in reasonable
agreement with typical magnetic fields 
inferred in laser-plasma experiments \citep{NilsonPRL2006, PetrassoPRL2009}.

\begin{figure*}
\centering
\includegraphics[width=\linewidth]{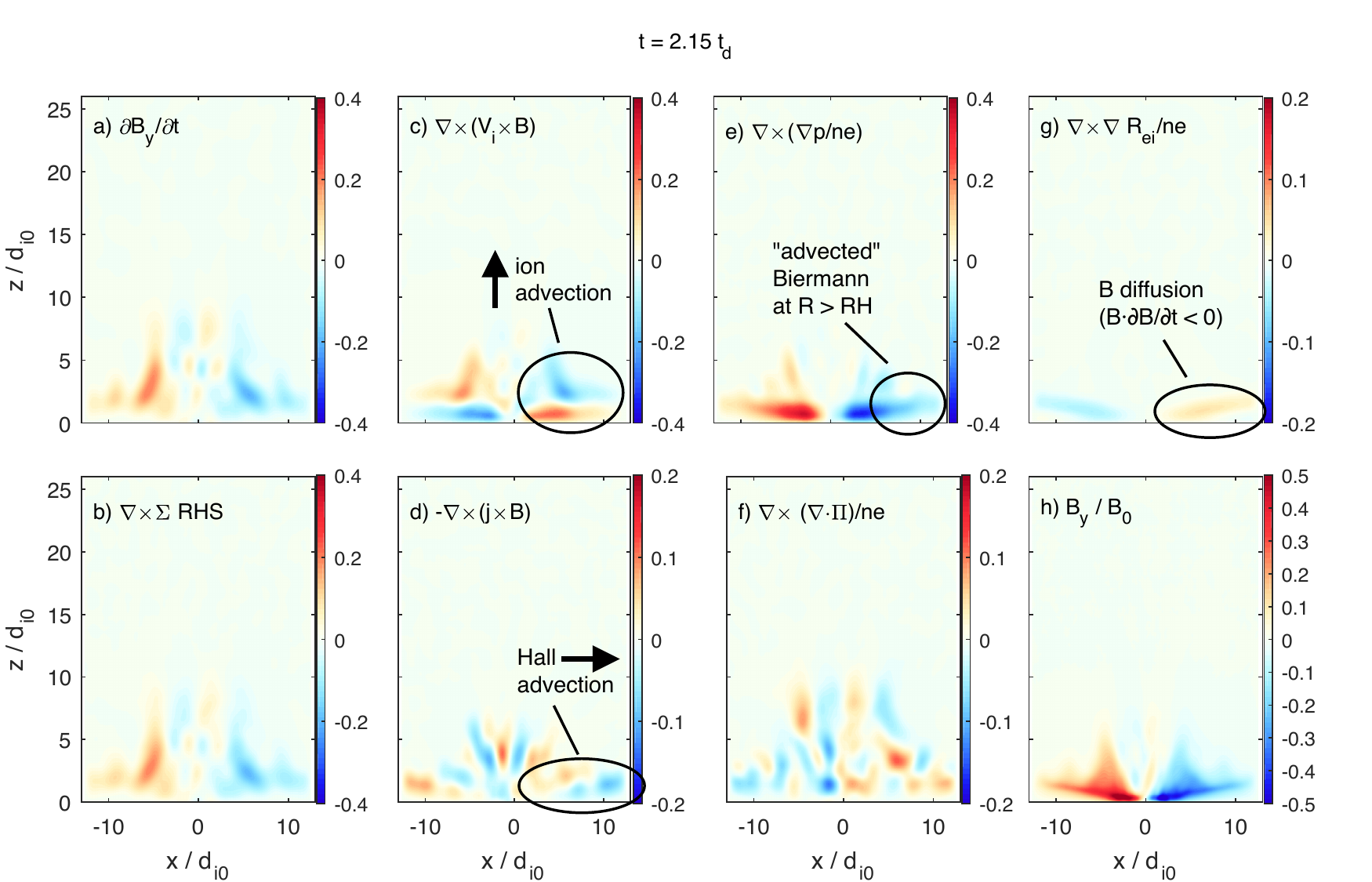}
\caption{Analysis of all terms of the Ohm's law for the magnetic field evolution in
simulations of the Vulcan experiments \citep{NilsonPRL2006}. \label{Fig_Ohm}}
\end{figure*}

The particle-in-cell simulation allows the full dynamics of
the magnetic fields to be tracked using the generalized Ohm's law.
Figure~\ref{Fig_Ohm} shows plots of all terms of the Ohm's law
versus space at the characteristic time $t=2.15 t_d$
when the magnetic field has expanded a significant distance
over the target surface.
The top-left panel shows $\partial B_y / \partial t$
(with units  $B_0 / t_d$) directly observed from the field evolution
in the simulation, whereas the panel
below it shows the \textit{inferred} evolution by taking the
curl of the measured Ohm's law, summing all measured effects.
These measured effects are broken down into:
the Biermann term; 
ion advection (curl $\mathbf{V} \times \mathbf{B}$); 
the Hall term (curl (1/ne) $\mathbf{j \times B}$); 
collisional momentum exchange ($\mathbf{R_{ei}}$); and 
the pressure tensor.
These terms are summed to produce $\mathbf{E_{ohm}}$, the curl of
which is plotted in the bottom left panel.  The excellent
agreement between the observed field evolution
and curl of Ohm's law indicates that all field
evolution effects are accounted for.  We now briefly
describe the role of these terms.

First, at this time the Biermann term continues to operate and is
dominant near the regions of strong radial temperature gradient,
near $x = \pm 5 d_{i0}$ (Fig.~\ref{Fig_Ohm}e).  The hotspot has also spread due to thermal transport,
leading to ``advected'' Biermann generation outside of the initial
heating radius ($R_H = 2 d_{i0}$).  We also quantify magnetic field
advection by multiple effects.
We observe a strong outward-radial advection from the Hall term (Fig.~\ref{Fig_Ohm}d),
which has a characteristic ``advection'' pattern $\partial B / \partial t \sim \nabla \times (v \times B)$; 
for example, for $x > 0$, we observe $\dot{B_y} > 0$ at $x= 5\,d_{i0}$ and $\dot{B_y} < 0$ at $x = +10\,d_{i0}$, 
which transports the negative $B_y$ outward.
This shows a fast radial advection due to the Hall
effect, an effect which has not previously been documented
to drive a fast radial expansion.
The Hall effect results from the differing electron and ion
flows associated with magnetic fields with gradients
on the local ion-skin-depth scale near the target surface.
Meanwhile, an advection pattern is 
also observed due to ion flow (Fig.~\ref{Fig_Ohm}c), however it 
predominantly transports the
magnetic field vertically off the target surface.
The simulation also tracks the momentum
exchange term.  This term contains both
advection from the Nernst term (also known as the thermal force) and collisional
diffusion.  However, we note that in this simulation, the
pattern is consistent with diffusion and
destruction of magnetic field rather 
than advection; that is, the signs of $R_{ei}$
are such to locally decrease the magnetic field everywhere, 
rather than bipolar signatures consistent with transport.

Finally, we note that due to finite collisionality 
in the simulation, we do not observe
an electron Weibel instability, as has been observed in collisionless
simulations in the same large $L/d_e$ regime \citep{SchoefflerPRL2014}.  However, we do
observe \textit{ion}-Weibel instability \citep{FoxPRL2013} due
to interaction of the expanding plasma
with the background plasma.
In this simulation, at relatively small $R_H/d_{i0}$, a single Weibel filament grows in this
simulation at the core of the bubble for $|x| < 2 d_{i0}$,
at $z = 5 d_{i0}$.  A trace of Weibel instability in the
expanding plasma is reflected in the alternating pattern of the
electron pressure tensor (Fig.~\ref{Fig_Ohm}f) near $z = +5 d_{i0}$.

\begin{figure*}
\centering
\includegraphics[width=\linewidth]{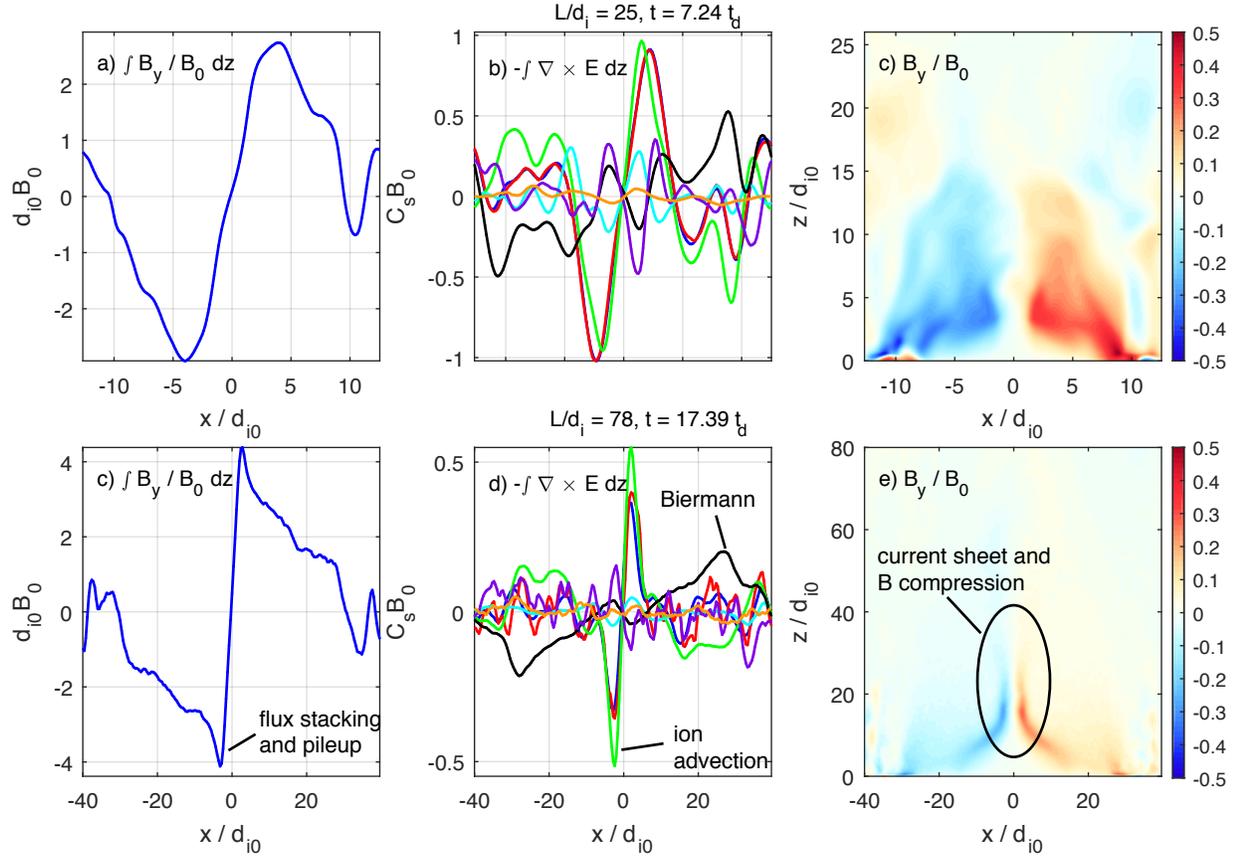}
\caption{Analysis of the evolution of magnetic fields in the simulations of
Vulcan \citep{NilsonPRL2006} and SG-II \citep{ZhongNature2010, DongPRL2012} experiments.
(a,d) Line-integrated magnetic fields, (b,e) vertically-integrated curl of Ohm's law
showing the contributions to magnetic flux evolution. (c,f) 2-D magnetic field
profiles showing current sheet formation when two plumes collide.
  \label{Fig_comparison}}
\end{figure*}

Simulations of colliding plasmas are conducted by moving the boundary
so that two plumes are centered, respectively, at the boundaries $x = \pm L$, which then
evolve and collide at $x = 0$.  A current
sheet forms as the opposite fields from the two plasmas interact
and compress, showing from first principles how and where a current sheet
forms for Biermann-battery based magnetic reconnection experiments.
Figure~\ref{Fig_comparison} shows current sheet formation
process and analysis by the Generalized Ohm's law.
The 2-D magnetic field profiles at the 
characteristic time of current sheet formation for both Vulcan (Fig.~\ref{Fig_comparison}c) and
SG-II experiments (Fig.~\ref{Fig_comparison}e), provides a comparison 
of the magnetic field produced for the two systems,
which differ primarily in the system size parameters $L / d_{i0}$ 
and $R_H / d_{i0}$.
For Vulcan, compressed magnetic fields in the current
sheet $B_y \approx 0.3 B_0$ are obtained through much of the plume,  whereas
for SG-II typical magnetic fields in the plume itself are near 0.1~$B_0$, but
there is a flux compression up by a factor $\approx 3$ in the collision
region leading to peak fields again near 0.3~$B_0$.
The larger $B$ field in the Vulcan case (scaled to $B_0$, though not in physical units) is
due to the smaller $(L,R_H) / d_{i0}$, consistent with
typical Biermann scaling \citep{SchoefflerPRL2014}.

Figure~\ref{Fig_comparison}(a,c) show quantitative comparison of the line-integrated
magnetic field $\int \mathbf{B} \times d\ell = \int B_y \,dz$, which
is useful for comparison with experimental
proton-radiography measurements \citep{PetrassoPRL2009,RosenbergPRL2015},
which produce maps of the line-integrated fields.
We obtain $\int B_y dz$ in the range of 1--3~$B_0 d_{i0}$
for the two experiments, which corresponds to 
experimental values in the range of 5--15~T-mm.  
In the case 
of the SG-II experiments, we note the simulation observes
a flux compression and stacking into the current layer,
such that $\int B_y dz$ peaks by about 30\% in
the reversal region compared to immediately upstream.
This demonstrates that the magnetic field is
significantly modifying the hydrodynamic evolution.

Figure~\ref{Fig_comparison}(b,d) analyzes the terms
driving the evolution of $\int B_y dz$, namely that 
the evolution of line-integrated magnetic field is driven by
the line-integrated Ohm's law,
$(\partial/\partial t) \int B_y dz = \int (\nabla \times E)_y dz$.
This shows which
terms are playing the greatest role on average
in evolving the magnetic flux,
which is valuable given the richness of physics
demonstrated in Fig.~\ref{Fig_Ohm}.
First, the red and blue curves are the directly measured
$\partial B_y/ \partial t$ and summed terms of 
Ohm's law, demonstrating excellent agreement and that
all effects have been measured.  The bi-polar enchancement of
$\partial B_y/ \partial t$ near $x = 0$ is
consistent with a current sheet that is still compressing and thinning.
The black curves show the Biermann term,
which peak in the Biermann-generation regions in the 
core of the bubbles near regions of
large temperature gradients.  (Again, the discussion here
is for vertically-integrated quantities; for brevity we will not mention this each time.)
We note there is a greater spatial separation 
between Biermann generation region and the current
sheet in the SG-II case, because the
greater $L / d_{i0}$.
The green curve shows the ion flow, which dominates in advecting
magnetic field away from the Biermann formation regions and
into the forming current sheet.
The light-blue curve is the Hall term which plays a role to advect 
the magnetic field into a narrower current sheet than the ion term alone.
The purple curve shows magnetic field destruction by
the off-diagonal pressure term.
The orange term shows a small amount of magnetic field destruction
by collisional diffusion, which is also largely negligible.

The simulations here demonstrate how the 
flows generate field and advect it toward the collision region, 
showing for the first time the details of the formation 
of a current sheet where $\textbf{B}$ reverses over a narrow
layer.  A follow-up paper will analyze 
the subsequent magnetic reconnection, which requires full 3-D simulations 
to allow reconnection outflows \citep{Matteucci2017}.

\section{Formation of magnetized collisionless shocks}
\label{SectionShock}

Shockwaves in space and astrophysical plasmas occur where plasmas
interact at super-sonic or super-magnetosonic velocity
\citep{SmithScience1975, SmithScience1980, BurlagaNature2008, SulaimanPRL2015}.
In collisionless plasmas the
shock layer forms at kinetic plasma scales through
collective electromagnetic effects, which convert the flow energy to heat 
and accelerate 
 non-thermal particle populations, \citep{BambaApJ2003,MastersNature2013}, 
 including cosmic rays \citep{KazanasNature1986, LoebNature2000, AckermannScience2013}.
Of particular importance are the class of supercritical shocks \citep{Balogh2013} (M $\gtrsim 3$), which
propagate faster than can be accommodated by entropy production in the shock, and therefore
reflect significant numbers of particles back into the upstream, seeding 
proposed cosmic ray acceleration processes \citep{BlandfordApJ1978, CaprioliApJ2014, MatsumotoScience2015, McClementsPRL2001} such as diffusive
shock acceleration (DSA).  Spacecraft routinely observe shocks
both at planetary bow shocks, and embedded in
solar wind flows.  
Recent spacecraft such as MMS 
produce detailed high-resolution data sets of collisionless shock
crossings  \citep{JohlanderPRL2016}.  
Nonetheless, controlled, repeatable laboratory experiments offer opportunities to
study aspects which are challenging with single- or few-spacecraft shock crossings,
including higher-dimensional and temporal effects such as rippling instabilities
and shock reformation, and particle energization by shocks.

Recent laboratory experiments 
have demonstrated the generation of shocks in magnetized plasmas driven by 
laser-driven pistons both in a low-Mach-number ($M_A \sim 2$) regime  
\citep{NiemannGRL2014, SchaefferPoP2015},
and in a high-Mach-number regime ($M_A \sim 15$) \citep{SchaefferPRL2017}.
We briefly review the opportunities for laboratory experiments.
First, laboratory experiments may allow investigation of 
higher dimensional stability and dynamics of shocks beyond
standard 1-D quasi-steady shock theories.  
One manifestation of this is the reformation
process of the shock, where the ``foot'' of reflected ion gradually grows until it disrupts the 
shock, leading to formation of a new shock at the location of the
erstwhile ion foot.  
Non-stationarity has been observed in 1-D plasma wind tunnel experiments \citep{MorsePRL1972}.
A related process is transverse
rippling of the shock front by instabilities \citep{LoweAnnGeo2003, UmedaApJ2009,YangJGR2012}.
At extreme mach number ($M_A \sim 40$) the front becomes 
turbulent by counterstreaming Weibel instability \citep{MatsumotoScience2015},
which is very interesting as these instabilities may be help 
energize particles near the shock front and inject them into DSA \citep{UmedaApJ2009, YangJGR2012}.
While some of these processes have been documented recently 
by spacecraft \citep{LobzinGRL2007,JohlanderPRL2016}, in general this is
difficult with single- or few-spacecraft crossing but it is readily
possible with imaging diagnostics available in laboratory experiments.

Second, shock acceleration is believed to be very important 
in astrophysical plasmas, and there is strong evidence for
particle acceleration near to shocks---
for example the TeV synchrotron emission \citep{BambaApJ2003}
and pion decay signatures conclusively
demonstrate electrons and cosmic ray protons are accelerated in 
blastwaves of supernova remnants \citep{AckermannScience2013}.  However, very little can
be constrained about the shock environment and acceleration process 
from these remote sensing observations.
While local spacecraft observations have demonstrated particle acceleration near shocks \citep{MastersNature2013},
laboratory experiments can contribute significant understanding of shock acceleration 
through parametric studies of
the dependence of acceleration mechanisms on plasma and shock parameters.
Of particular importance is the physics and efficiency of the ``injection'' process
by which particles obtain a pre-acceleration which enables them to 
then participate in the DSA.

A recent series of experiments have developed a new
platform to study the physics of 
high Mach-number flows, including magnetized shocks and reconnection, 
in the laboratory \citep{FikselPRL2014,SchaefferPRL2017}.
These experiments, 
conducted at the OMEGA EP facility 
at the University of Rochester Laboratory for Laser Energetics,
demonstrated how to generate controlled 
supersonic ablation flows into a pre-magnetized ambient plasma.
The hallmark of the platform is that 
externally-controlled magnetic fields are 
generated by a compact pulsed power system \citep{GotchevRSI2009}, 
generating field of order 10~T over a volume several-mm on a side,
which is used to create a pre-magnetized ambient
plasma as a medium for magnetized shocks and flows.

In this section we extend the kinetic ablation simulations
of the previous sections to study plasma expansion into a magnetized 
background plasma to model this experimental system.
Kinetic simulations played an essential role in
developing this experimental platform by providing intuition on the
behavior of the system over broad parameter regimes, 
and by pointing out experimental observables.
Recently published simulations with this model have shown 
how the expanding plasma sweeps up the ambient plasma and field,
leading to flow and density steepening and magnetic compression,
the onset of ion-reflection, and finally the 
formation of a magnetized high-Mach number shock \citep{SchaefferPRL2017}.
A slight variant of the model, which using a particle
source term rather than a heating operator, was used 
to simulate the reconnection experiments in Ref.~\citep{FikselPRL2014},
which showed important features of the experiments
including flow stagnation of two colliding plasmas
at the ion-skin-depth scale, and current sheet tearing during the 
strongly-driven magnetic reconnection.
Here we describe the computational model and parameters in detail 
and show new results
of these processes obtained with large-scale 2-D shock simulations.

\subsection{System setup and parameters}

\begin{table}
\centering
\caption{Parameters for OMEGA-EP magnetized plume experiments}
\begin{tabular}{lr}
\tableline\tableline

Input: \\
\tsp $n_{ab}$ (m$^{-3}$, DRACO) & \sciexp{6}{26} \\
\tsp $n_{bg}$ (m$^{-3}$) & \sciexp{6}{24} \\
\tsp $T_{ab}$ (eV, DRACO) & 800 \\
\tsp heating radius $R_H$  ($\mu$m)& 350  \\
\tsp heating exponent $\kappa$ & 4  \\

\\

Dimensional: \\
\tsp $d_{e0}$ ($\mu$m) & 0.24  \\
\tsp $d_{i0,CH}$ ($\mu$m) & 14  \\
\tsp $C_{s0}$ (m/s) & \sciexp{2}{5} \\
\tsp $B_0 = (\mu_0 n_{ab} T_{ab})^{1/2}$ (T) & 300 \\
\tsp $B_{up}$ (T) & 8 \\
\tsp $t_d = d_{i0}/C_{s0}$ (ps) & 52  \\
\tsp $\omega_{cH,up}^{-1}$ (ns) & 1.3 \\
\tsp $\omega_{cC,up}^{-1}$ (ns) & 2.6 \\

\tsp $\omega_{ce0}$ (ps$^{-1}$) & 50 \\
\tsp $\nu_{ei0}$ (ps$^{-1}$) & 4 \\
\\
Dimensionless: \\

\tsp $R_H / d_{i0}$ & 30 \\
\tsp $\lambda_{\mfp 0} / d_{e0}$ & 12  \\
\tsp $B_{up} / B_0$ & .025 \\
\tsp $n_{bg} / n_{ab}$ & .01 \\

\end{tabular}
\label{TableShockExp}
\end{table}

DRACO radiation-hydrodynamics simulations are again
used to provide plasma parameters to drive the PSC ablation simulations.
Table~\ref{TableShockExp} lists the 
relevant ablation parameters predicted by DRACO,
which used the laser geometry and parameters of the experiments \citep{FikselPRL2014, SchaefferPRL2017},
including the highly oblique laser-incidence, $74^\circ$,
and the 
EP 750-$\mu$m distributed phase plate, which generates a smooth beam profile
with $1/e$ radius $R_H = 340~\mu$m and a 4th-order
super-gaussian profile (laser intensity $I \propto \exp[- (r / R_H)^\kappa]$,
with $\kappa = 4$).
We take an upstream magnetic field
near 8~T, and an upstream density of \sciexp{6}{24}~m$^{-3}$, which is somewhat above typical
DRACO predictions 
and recent measurements, but was used as part of a scan to confirm shock formation for
a variety of upstream densities.
The plasma ablated from the targets is a 60-40 mixture of 
hydrogen and carbon  (C$^{6+}$) for which DRACO uses an average ion model.

For the present paper, we consider an ``equivalent'' kinetic PSC 
simulation based on a single species.  By this we mean that 
the ion-scale parameters are chosen to match Table~\ref{TableShockExp},
but are used in a simulation with only H+ ions, for
simplicity.  Ablation timescales ($t/t_d, t \omega_{ci,up}$) and length scales ($x/d_{i0}$)
are measured in the pure PSC simulation in relation to this lone species.
Extension to multiple species in the kinetic simulation is straightforward, and in fact was 
used in \citet{SchaefferPRL2017} for accurate comparison with the experiment,
and will be considered in more detail in future work.
Parameters for the present simulations are presented in Table~\ref{TableShockSim}.
We note that in addition to the single ion species, we use slightly increased
magnetic fields $B_{up}/B_0$ to obtain faster shock formation
in a given simulation wall time, and we use lower electron collisionality 
$\lambda_{\mfp 0} / d_{e0} \approx 50$ than estimated for the experiments.
The plasma ablation flows in this PSC simulation and DRACO are compared in Fig.~\ref{Fig_draco_psc},
where the good agreement of the two flows supports the scaling via the ion-scale parameters between 
these two very different simulation codes.

\begin{table}
\centering
\caption{PSC Shock Simulations}
\begin{tabular}{lr}

\tableline \tableline

Input:\\
\tsp $R_H / d_{i0}$ & 30 \\
\tsp $\lambda_{\mfp 0} / d_{e0}$ & 50  \\
\tsp $B_{up} / B_0$ & 0.04 \\
\tsp $n_{bg} / n_{ab}$ & 0.01 \\

Electron-scale parameters:\\
\tsp $M_i / m_e$ & 100 \\
\tsp $m_e c^2 / T_{e,ab}$ & 16 \\

Results: \\
\tsp $t_{shock} \omega_{ci,up} $ & $\sim 1$ \\
\tsp $V_{shock} / C_{s0}$ & 4.5 \\

\end{tabular}
\label{TableShockSim}
\end{table}

\subsection{Results}

\begin{figure*}
\includegraphics{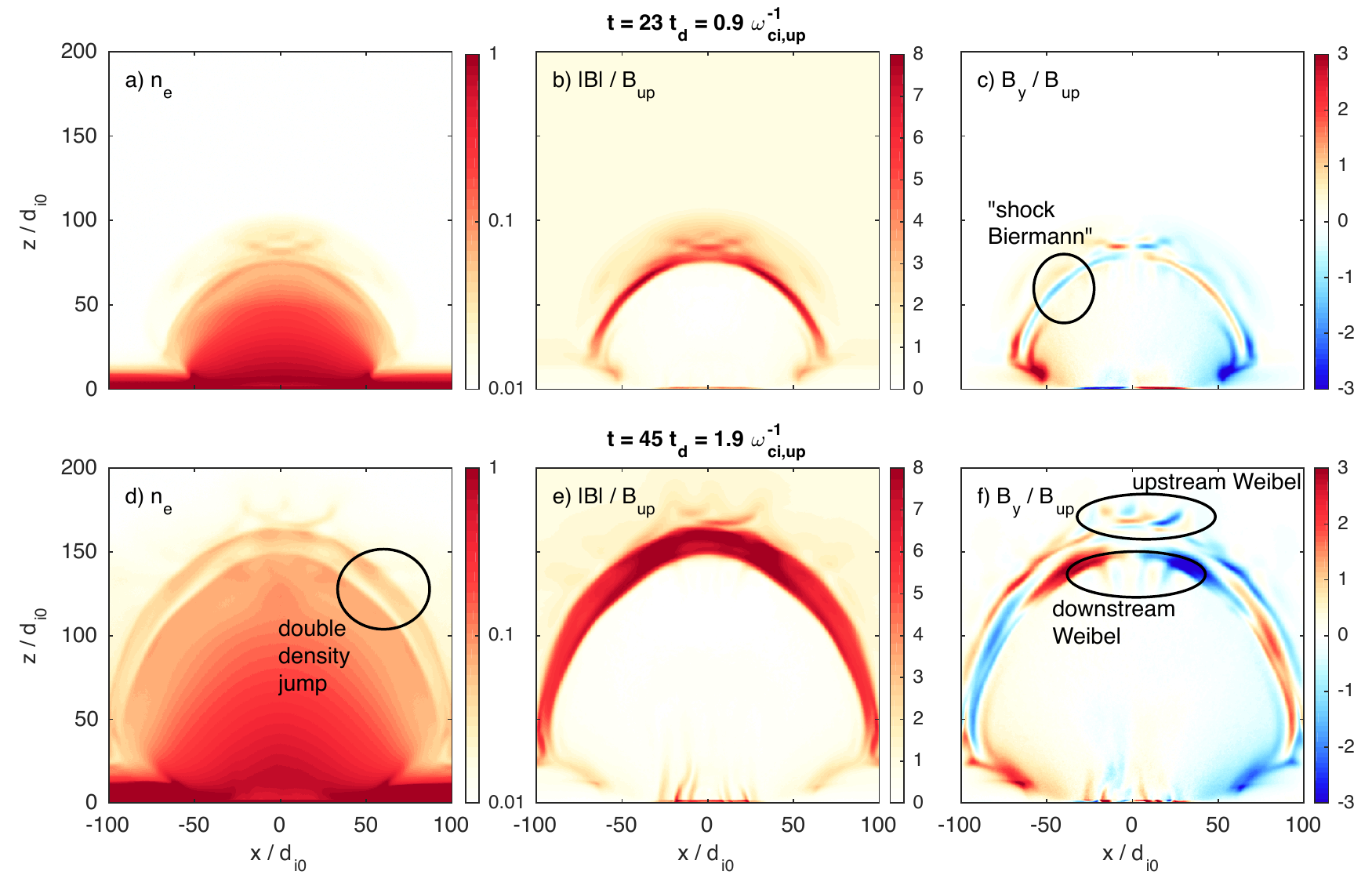}
\caption{Generation of magnetized shocks by ablation flow into a magnetized ambient plasma, at
two times characteristic of shock formation, $t \omega_{ci,up} = 0.9$ (a-c) and $t \omega_{ci,up} = 1.9$ (d-f).
(a,d) 2-D density maps; (b,e) total magnetic field $|B|$ emphasizing the compression of the initial upstream field
and formation of the diamagnetic cavity. (c,f) shows the additional out-of-plane field generation $B_y$
in the 2-D expansion by the Biermann effect, shock-Biermann, and Weibel instability.}
\label{Fig_psc_shock}
\end{figure*}

Figure~\ref{Fig_psc_shock} shows the results of large-scale 2-D simulations
on the expansion into the background plasma and generation of the 
magnetized shock.  The upper (a-c) and lower (d-f) panels show the 
evolution at two characteristic times, near the time of initial shock formation ($t \omega_{ci,up} = 0.9$, a-c)
and later as the shock completely separates from the piston ($t \omega_{ci,up} = 1.9$, d-f).
We scale the  time evolution in units of $t \omega_{ci,up}$ based on the initial upstream ambient field, 
as we have found that this is the 
most relevant time-scale for shock formation for a wide range of $t_d \omega_{ci,up} \sim \beta^{-1/2}$.
The left panels (a,d) show the plasma density evolution, which show the density compression
that forms as the ablation plasma flows into and compresses the ambient plasma, initially
 at $n_{bg} = 0.01 n_{ab}$.  The interaction of the two 
plasmas occurs where the exponential ablation plume and background plasma density
approximately coincide.  Because of the ambient magnetic field, the plasmas cannot fully
interpenetrate, and instead the field and plasma are swept up,
and steepen into a shock.  At late times (Fig.~\ref{Fig_psc_shock}d), the 
shock ramp separates from the piston, forming a characteristic ``double-jump'' of the density. 
This feature was in fact predicted first from PSC simulations and was used as 
an experimental observable of the shock formation in experiment, where 
the transition over time from single to double-density jumps 
are observed with interferometry \citep{SchaefferPRL2017}.  This shows
the value of kinetic simulations in interpreting experiments.

Figure~\ref{Fig_psc_shock}(b,e)  show the magnetic field evolution
in the simulations.  Initially the upstream magnetic field lies in the simulation
plane and is a uniform field of magnitude $B_{up}$.  However, this field is
swept up and compressed into the ablation flow.  The in-plane magnetic flux is conserved,
such that the magnetic field compressed into the shock is exactly
that expelled from the diamagnetic cavity near the targets.  Over time, 
the amount and magnitude of the magnetic expulsion increases, and these ablation-driven 
shocks continue to expand through the whole simulation.  
The magnetic field compresses to a value of about 6 $B_{up}$ by $t \omega_{ci,up}$ = 1.9.

Meanwhile, Fig.~\ref{Fig_psc_shock}(c,f) shows additional magnetic field
\textit{generated} near the shock by the plasma.
This magnetic component is the out-of-plane (azimuthal in a toroidal geometry)
and is not present in the initial condition, but as shown in the 
previous section is readily generated by a variety of processes.
We note first of all the Biermann field generation near the target, which is analogous to
the pure Biermann generation in unmagnetized expansions shown previously in Fig.~\ref{Fig_evol}.
Perhaps unsurprisingly, 
the Biermann field is also swept up by the plasma and
compressed into the shock.  However, we note that the Biermann field is
far from uniform in polarity, and shows regions of both signs depending on the polar location
within the shock.  Fields of the same sign as the standard Biermann fields are
generated in equatorial regions (near the target surface), 
but regions with opposite polarity are generated at oblique polar angles.  We refer to these
as ``shock Biermann" fields, as this is kinetic Biermann field generation 
within a collisionless shock.  We believe this is the first evidence for this process
which should be considered more broadly in astrophysical and laboratory shocks.
Furthermore, the field quantitatively modifies the magnetic fields 
in the shock, with $B_{\textrm Biermann} \sim 3 B_{up}$ reaching
approximately 50\% of the compressed shock field $B_{\textrm shock} \sim 6 B_{up}$ under the 
present parameters.  The Biermann fields within the shock can quantitatively 
(or even qualitatively, if the upstream field is weak enough)
modify shock formation via enhancing particle reflection.  This topic is 
the subject of ongoing studies.

Finally, near the poles, the simulations also show the development of out-of-plane
filamentary fields characteristic of Weibel instability \citep{FoxPRL2013}, 
both upstream and downstream of the shock. 
The upstream Weibel is a transverse instability of the reflected ions 
kicked forward into the upstream plasma, whereas the downstream Weibel results from 
ablated ion gyration off the compressed fields, returning into the downstream.
Weibel instability has been proposed to enhance
 particle energization in high-Mach number shocks \citep{MatsumotoScience2015}.
These simulations therefore show the interaction of this instability with 
scaled laboratory shocks, and suggest that both these processes may be 
observable in laboratory experiments.

\section{Discussion and conclusions}

In this paper, we have developed fully kinetic simulation techniques
to model recent laser-plasma experiments on laboratory astrophysics,
allowing simulation of a variety of 
phenomena relevant to astrophysical plasmas
within a common framework.

First, we set out an overall methodology based on 
dimensionless parameters for designing scaled simulations to
model these processes.   
Ion-scale dimensionless parameters are derived 
and provide a basis to initialize the simulations with
relevant parameters to match experiments.
This is important because for fully kinetic simulations over
these plasma length scales ($L/d_i \sim 100$), evolving for multiple
sound crossing times $(t C_s / L \gtrsim 1$), it
is still not computationally practical to match
all the dimensionless parameters of the system,
especially those involving the small electron mass ($M_i/Z m_e$), and 
the speed of light ($m_e c^2 / T_e$).
We note that kinetic simulations with reduced parameters 
like those shown here are very common in dedicated
simulations of 
magnetic reconnection \citep{BirnJGR2001, FoxPRL2011, FoxPoP2017} 
and shocks \citep{KatoApJL2008, MatsumotoScience2015}; 
here this technique is extended to general ablation flows for this
whole family of phenomena.  
These dimensionless parameters also provide a basis to
scale the results to astrophysical plasmas.
Some limitations of the present model include the fact that the
target does not reach up to solid densities, and the
lack of a laser-ray-trace energy-deposition model which requires input
from a separate radiation-hydrodynamics simulation.  Both
of these limitations may be improved by further code development.
However, already the model is in general agreement
with the expanding plasma evolution of DRACO.

Subsequently, we showed how the kinetic ablation model is used to 
simulate two sets of recent laser-driven laboratory astrophysics experiments.
First, we consider the Biermann battery magnetic field generation in recent
experiments with single and multiple expanding plumes.
The simulations results are directly analyzed via the generalized
Ohm's law and show both the location and time of Biermann-battery
field generation, and the mechanisms (ion flows, Hall flows) which advect these fields
and form a current sheet when two neighboring plumes collide.
The compression of the 
field indicates that the magnetic field has 
modified the hydrodynamic evolution of the two plasmas.
The 2-D simulations shown here do not include the magnetic reconnection
phase of the experiments, which requires a 3-D simulation,
but such simulations are underway and will be reported in 
a separate publication.  The results provide insights into the 
evolution of ablated plumes of wide applicability in 
laboratory plasmas and provide insights for designing
experiments and interpreting results on magnetic reconnection,
including rates of magnetic reconnection and efficiency
of acceleration of particles.

Finally, we extended the simulation to study plasma expansion into
a pre-magnetized background plasma, modeling recent experiments
on magnetic reconnection and collisionless shock formation.
The simulations provide important insights for 
interpreting the experiments, including confirmation of the formation of
shocks on the timescale $\omega_{ci,up}^{-1}$.
These large-scale 2-D simulations also show the development
of Biermann battery magnetic fields within the 
collisionless shock front, which quantitatively
modify the (compressed) shock fields, as
Biermann fields in the shock up to about 50\% of the compressed
shock fields are generated.
We note that the development of Biermann fields in
collisional, MHD-scale shocks was considered
recently in Ref.~\citep{GrazianiApJ2015},
which presented analytic arguments for the 
magnitude and sign of the magnetic field, with
proposed prescriptions of how to model them in reduced fluid models.  A comparison
between kinetic and MHD simulations 
is potentially of great value in benchmarking these reduced models.
The simulations also show the development of 
transverse Weibel instability both ahead and behind the
shock, driven in the first case by the typical 
forward particle reflection off the shock at high Mach-number, and 
behind the shock by backward gyration or rebound
of the piston ions off the magnetic compression.
The results provide insights which will help design future experiments 
on transverse shock stability and particle acceleration
by shocks.

\acknowledgements 

Simulations were conducted on the Titan supercomputer at the Oak Ridge Leadership Computing 
Facility at the Oak Ridge National Laboratory through the 
Innovative and Novel Computational Impact on Theory and Experiment (INCITE) program, which is
supported by the Office of Science of the DOE under Contract No. DE-AC05-00OR22725.
This research was also supported by the DOE under Contracts No. DE-SC0008655 and No. DE-SC0016249.


%

\end{document}